\documentclass[letter]{aa}

\usepackage{natbib}
\bibpunct{(}{)}{;}{a}{}{,}

\usepackage{graphicx}
\usepackage{txfonts}
\begin{document}
\def\gapprox{\;\rlap{\lower 2.5pt            
 \hbox{$\sim$}}\raise 1.5pt\hbox{$>$}\;}       
\def\lapprox{\;\rlap{\lower 2.5pt            
 \hbox{$\sim$}}\raise 1.5pt\hbox{$<$}\;} 
%
%
\title{3D simulations of RS~Oph: from accretion to nova
  blast\thanks{In the frame of the computing project 'Cosmic Engines in Galaxies'}}

\author{Rolf Walder\inst{1,2,3}
        \and
        Doris Folini\inst{2}
        \and
        Steven N. Shore\inst{3,4}
      }

\offprints{Rolf Walder}

\institute{Observatoire de Gen\`eve, 
  Universit\'e de Gen\`eve, 
  51 ch. des Maillettes, 1290 Sauvernay,   Switzerland
  \email{walder@astro.phys.ethz.ch}
  \and
  Institute for Astronomy, ETH Z\"urich,
  ETH-Zentrum, SEC D2, CH-8092 Z\"urich, Switzerland \\
  \email{folini@astro.phys.ethz.ch}
  \and
  Dipartimento di Fisica 'Enrico Fermi',
  Universit\`A di Pisa, largo Pontecorvo 3, Pisa 56127, Italy
  \email{shore@df.unipi.it}
  \and
  INFN/Pisa, largo Pontecorvo 3, Pisa 56127, Italy
}

\date{Received ; accepted}

\abstract{RS~Ophiuchi is a recurrent nova with a period of about 22
  years, timescale, consisting of a wind accreting binary system with
  a white dwarf (WD) very close to the Chandrasekhar limit and a red
  giant star (RG). The system is considered a prime candidate to
  evolve into an SNIa. For its most recent outbursts in 1985 and 2006,
  exquisite multiwavelength observational data are available.}
{Further physical insight is needed regarding the
 inter-outburst accretion phase
 and the dynamical effects of the subsequent nova explosion in order
 to improve the interpretation of the observed data and to shed light
 on whether the system is an SNIa progenitor.}
{We present a three dimensional hydrodynamic simulation of the quiescent
  accretion and the subsequent explosive phase.}
{The computed circumstellar mass distribution in the quiescent phase
  is highly structured with a mass enhancement in the orbital plane of
  about a factor of 2 as compared to the poleward directions.  The
  simulated nova remnant evolves aspherically, propagating faster
  toward the poles. The shock velocities derived from the simulations
  are in agreement with those derived from observations. For
  $v_\mathrm{RG} = 20$~km/s and for nearly isothermal flows, we derive
  a mass transfer rate to the WD of 10\% of the mass loss of the RG.
  For an RG mass loss of $10^{-7} M_\mathrm{\odot}$yr$^{-1}$, we found
  the orbit of the system to decay by 3\% per million years.  With the
  derived mass transfer rate, multi-cycle nova models provide a
  qualitatively correct recurrence time, amplitude, and fastness of
  the nova.}
{Our 3D hydrodynamic simulations provide, along with the observations
  and nova models, the third ingredient for a deeper understanding of
  the recurrent novae of the RS~Oph type. In combination with recent
  multi-cycle nova models, our results suggests that the WD in RS~Oph
  will increase in mass. Several speculative outcomes then seem
  plausible. The WD may reach the Chandrasekhar limit and explode as
  an SN~Ia. Alternatively, the mass loss of the RG could result in a
  smaller Roch volume, a common envelope phase, and a narrow WD + WD
  system. Angular momentum loss due to graviational wave emission
  could trigger the merger of the two WDs and -- perhaps -- an SN~Ia
  via the double degenerate scenario.}
\keywords{stars: individual: RS Oph -- stars: novae, cataclysmic
  variables -- supernovae: general -- accretion, accretion disks --
  Gravitational waves -- methods: numerical}

\maketitle
\section{Introduction}
Type Ia supernovae (SNe Ia) are cornerstones of modern cosmology, as a
measure for cosmological distances, and they are crucial building
blocks of the universe, as production sites of a large part of iron
group elements. The recurrent nova RS Ophiuchi (RS Oph)
\citep{hachisu-et-al:99, 2001ApJ...558..323H} is a candidate for their
still mysterious progenitors. It is a symbiotic-like binary star
system consisting of a red giant (RG) and a white dwarf (WD) that
undergoes a nova outburst about every 22 years \citep{anupama:02,
  fekel-et-al:00, dobrzycka-et-al:96}. For the most recent outbursts
in 1985 and 2006, exquisite panchromatic observational data are
available \citep{shore-et-al:96, bode-et-al:06, sokoloski-et-al:06,
  das-et-al:06, obrien-et-al:07, 2006MNRAS.373L..75E,
  worters-et-al:07, hachisu-et-al:07, bode-et-al:07}.  Here we present
first 3D hydrodynamical simulations of RS~Oph, from pre-outburst
accretion through nova explosion, with the goal of improving the
physical insight into the system and the interpretation of the unique
observational data.

RS Oph has an orbital period of 455 days~\citep{fekel-et-al:00}, RG
and WD masses of 2.3 $M_\mathrm{\odot}$ and close to 1.4
$M_\mathrm{\odot}$ \citep{anupama:02, worters-et-al:07}, respectively,
and a separation between the components of $a = 2.68\cdot 10^{13}$~cm.
The RG mass is, however, still uncertain, much smaller
values~\citep{fekel-et-al:00} have also been suggested. We supplement
the system parameters by assuming for the RG a terminal wind velocity
of $v_\mathrm{RG} = 20$~km/s in the rest frame of the RG and a mass
loss rate of $10^{-7}$ $M_\mathrm{\odot}$yr$^{-1}$. Note that mass
losses of red giants are still not well known.
While~\citet{1990ApJ...349..313S} find a massloss of order $10^{-7}
M_\mathrm{\odot}$yr$^{-1}$ for the majority of RG in symbiotic
systems, the scatter is significant and values of less than $10^{-8}
M_\mathrm{\odot}$yr$^{-1}$ can be found.  For the range of spectral
types suggested~\citep{worters-et-al:07} for RGs in the RS Oph system,
its radius is always smaller than its Roche lobe and accretion by the
WD occurs only from the RG wind.
\begin{figure}[tp]
\centerline{\includegraphics[width=9.cm]{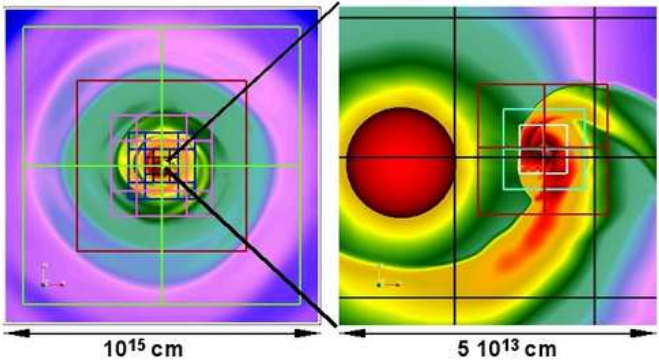}}
\caption{The grid structure of the nine levels of refinement, shown
  for the entire computational domain (left) and around the accreting
  WD (right). While the coarser grids of level 1 through 6 are fixed
  in space, the meshes of level 7 through 9 follow the WD. The RG is
  shown in red, the accreting WD in blue.}
\label{fig:density_mesh}
\end{figure}
\section{3D simulations of RS Oph}
\subsection{Computational method}
Our numerical simulation was performed using the
A-MAZE-code~\citep{walder-folini:00}, a parallel, block structured,
adaptive mesh refinement (AMR) hydrodynamical code using Cartesian
meshes and a multidimensional high-resolution finite-volume
integration scheme. The code has been used for accretion studies
before~\citep{zarinelli-et-al:95, walder:97, dumm-et-al:00,
  harper-et-al:05}. The Euler equations are solved in three dimensions
with a nearly isothermal polytropic equation of state with $\gamma
=1.1$, resulting in a thermal structure that comes close to that
obtained from photoionization models of related symbiotic binary
systems~\citep{1987A&A...182...51N, nussbaumer-walder:93}.

The simulation is carried out in an Eulerian frame of reference with
the stars moving within the computational domain which, which measures
$10^{15}$~cm a side. The computational grid consists of nine nested
levels of refinement (Fig.~\ref{fig:density_mesh}). The coarsest level
consists of 50 cells cubed. From one level to the next, grid cells are
refined by a factor of two. Each level comprises between 8 and 64
individual grids, the entire mesh consists of 233 grids and $2 \cdot
10^7$ cells.  The decomposed grid structure is exploited for
parallelization under OpenMP.

We impose free outflow at the outer domain boundary.  The WD is
represented as a spherical low pressure, low density, zero velocity
region of radius $\mathrm{R}= 2.2 \cdot 10^{11}$~cm. The accretion
rates of mass, momentum, energy, and angular momentum onto the WD are
derived from the flow quantities entering the region. The rates are
not sensitive to the size, pressure, and density of the region
representing the WD. The RG is represented as a spherical region of
150 $\mathrm{R_\mathrm{\odot}}$. In the rest frame of the RG star,
which rotates and orbits with respect to the computational domain, we
chose a terminal velocity of the RG wind of 20 km/s. The absolute
value of the density in the cells representing the RG was adjusted to
achieve the desired mass loss of $10^{-7}
\mathrm{M_\mathrm{\odot}}$yr$^{-1}$. The wind temperature is set to
8000~K, a value between the RG photospheric temperature and the wind
temperature more close to the WD as computed by elaborate
photo-ionization models~\citep{1987A&A...182...51N}.

At the beginning, the entire computational domain was filled with the
RG wind. The accreting system was then relaxed over seven orbital
periods, about the crossing time from the RG to the domain boundary at
the RG wind speed. Mass and angular momentum losses out of the system
were then computed through spherical shells around the centre of mass.

The losses became constant for shell radii larger than $10^{14}$~cm.
The nova explosion was then launched into the relaxed density structure.
\begin{figure}[tp]
\centerline{\includegraphics[width=9.cm]{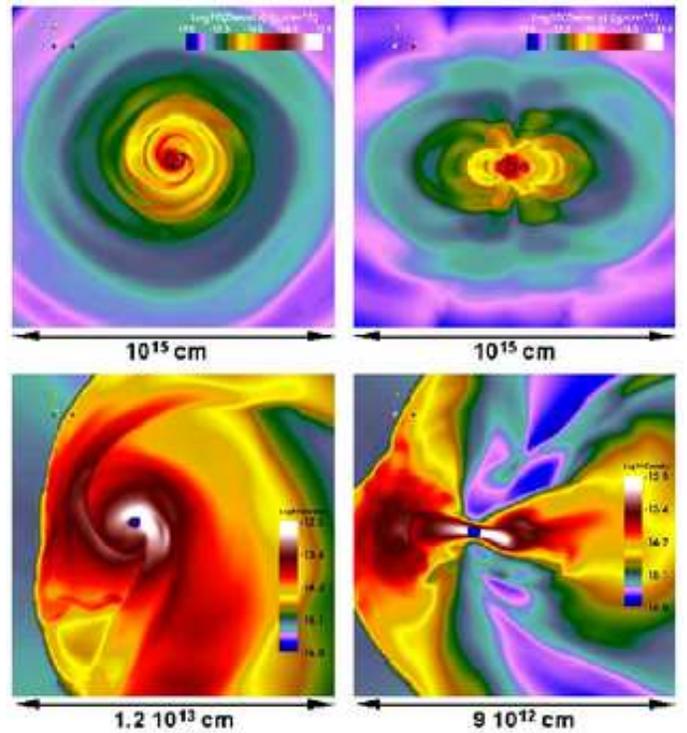}}
\caption{Density structure during the accretion phase. Shown is
  density (logarithmic scale, g/cm$^3$) in the orbital plane
  (xy-plane, left), and in a plane perpendicular to the orbit
  (yz-plane, right) for the entire computational domain ($10^{15}$~cm
  a side, top) and a zoom around the accreting WD ($\approx
  10^{13}$~cm a side, bottom). Self-interacting high-density spirals
  dominate the inner region up to a few times the binary separation.
  The RG is shown in red, the WD in blue.}
\label{fig:density_acc}
\end{figure}
\subsection{Quiescent phase: density spirals and wind accretion}
The inter-outburst accretion phase leads to a circumstellar
density distribution that deviates substantially from the $1/r^2$
density distribution of a single RG wind out to several system
separations.  The deviations, usually neglected when interpreting RS
Oph observations, arise from the orbital motion of the two components
and the accretion onto the WD, and are subsequently transported
outward in the flow. The resulting density structure,
Fig.~\ref{fig:density_acc}, depends critically on the ratio $R =
v_\mathrm{RG}/v_\mathrm{Orb},$ where $v_\mathrm{Orb} =2 \pi a / P$,
with $P$ the orbital period. For $R >> 1$, an Archimedian spiral
occurs in the orbital plane, as observed in the colliding wind Wolf
Rayet binary star system WR 104~\citep{1999Natur.398..487T}. For RS
Oph, with $R \le 1$, the spiral pattern becomes self-interacting.

In the vicinity of the WD, a high density accretion disk forms that is
clearly visible in Fig.~\ref{fig:density_acc}. 
Its diameter is roughly 1/10 of the system separation, and velocities
are non-Keplerian. Strong spiral shock waves are responsible for the
transport of angular momentum and mass, and stable accretion occurs.
The disk is geometrically thick, the opening angle perpendicular to
the orbital plane (Fig.~\ref{fig:density_acc}, lower right panel)
measures roughly $\pm 60^{\circ}$ in the direction toward the RG and
$\pm 30^{\circ}$ away from it.

\begin{figure}[tp]
\centerline{\includegraphics[width=9.8cm,height=4.8cm]{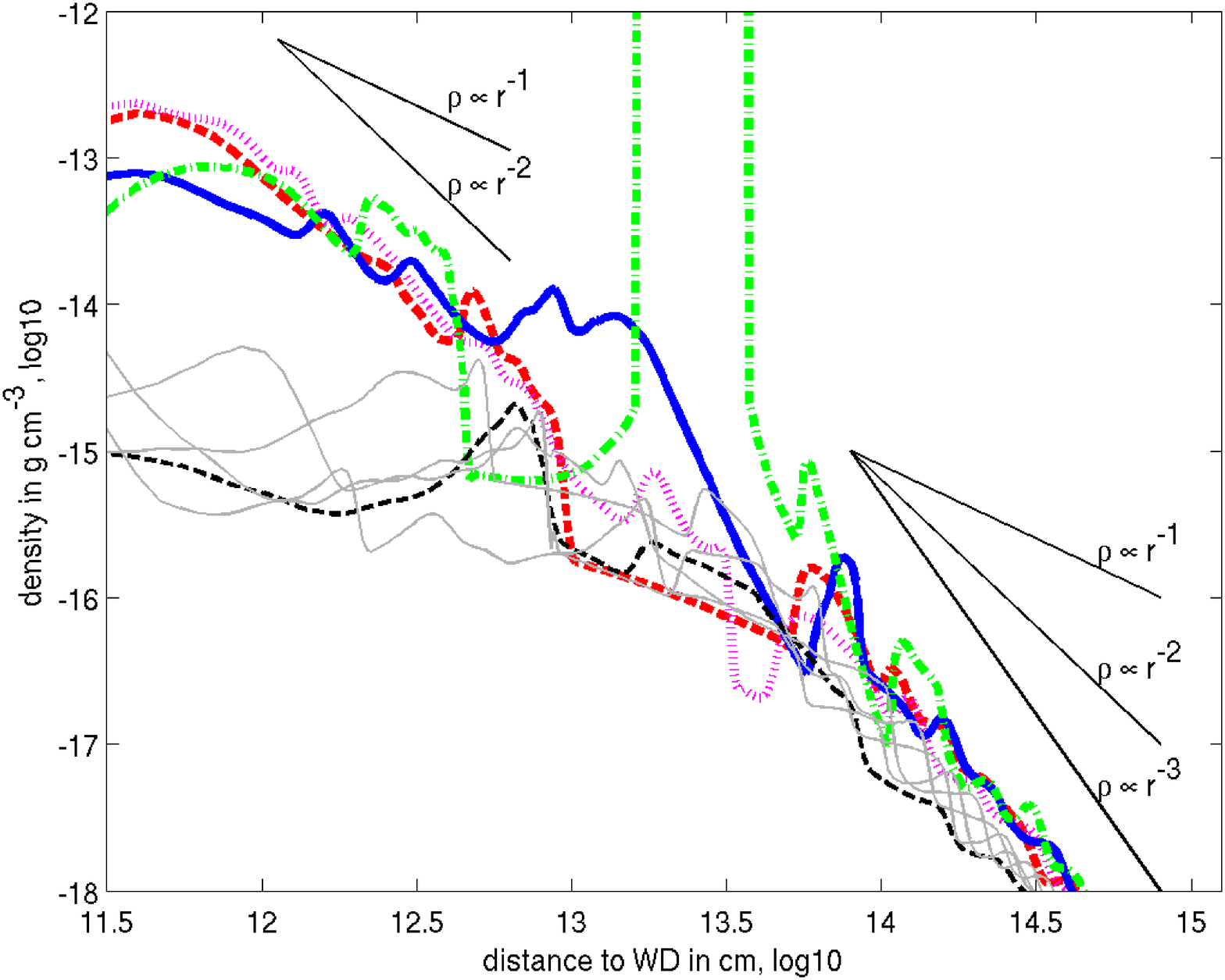}}
\caption{Density profiles along radial rays originating from the WD.
  Bold colored lines denote rays in the orbital plane along the x- and
  y-coordinate axis, the RG being located in the negative x-direction
  (-x green; +x magenta; -y blue; +y red).  Perpendicular, green bold
  lines denote the position of the RG.  The bold black line is in
  poleward direction ($+90^{\circ}$), thin grey lines belong to an
  inclination of $60^{\circ}$ and positive and negative x- and
  y-directions, respectively.}
\label{fig:densprofs_acc}
\end{figure}

On length scales less than the system separation, radial density
profiles as seen from the WD (Fig.~\ref{fig:densprofs_acc}) show
density contrasts between the orbital plane and poleward directions of
one to two orders of magnitude. Most prominent here, and reaching
furthest out, is the trailing accretion wake of the WD. This density
structure affects the early evolution of the nova remnant. At greater
distances, densities in the orbital plane remain larger than toward
the poles, but only by a factor of 2-3. This is, however, sufficient
to cause an asymmetric evolution of the nova remnant. Note that
although the density decreases as $1/r^2$ {\it on average}, the
relative amplitude of local variations due to the self-interacting
spiral pattern are up to a factor of 10, especially in the orbital
plane. We expect these local variations to vanish at even larger
scales that are beyond our computational domain.

The velocity of the systemic outflow is a superposition of the
velocity of the RG wind, the stellar orbital and rotational velocities
and hydrodynamical effects. On larger scales than $ \approx 5\cdot
10^{13}$~cm, the flow field is essentially directed radially outward.
Depending on the exact angle of an observer, any radial speed between
$20$~km/s (polewards) and $\approx 50$~km/s (mostly in the orbital
plane) can be found in the computational data.  This range is
consistent with the different values of the wind speeds derived from
observations, e.g. 36 km/s by~\citet{Iijima:07} and 40-60 km/s
by~\citet{shore-et-al:96}.

The {\it computed} accretion rate is 10 \% of the mass loss rate of
the RG, independent of the absolute mass loss of the RG. To our
knowledge, this is the first quantitative self-consistent prediction
of the accretion rate of a symbiotic-like recurrent nova.  A higher
accretion rate is expected for a less massive
RG~\citep{fekel-et-al:00, dobrzycka-kenyon:94}, since a smaller system
separation would result. The computed value is in line with the values
found in 3D simulations of related symbiotic binary star systems, for
example 6\% in RW Hydrae~\citep{dumm-et-al:00} and 10\% - 25\% in
Z~Andromeda~\citep{2005ARep...49..884M}. In absolute terms, the WD in
our simulation accretes $10^{-8} M_\mathrm{\odot}$yr$^{-1}$.  This
value also agrees well with the accretion rates required by
multi-cycle nova evolution models~\citep{yaron-et-al:05} for RS Oph
like binary systems, with $M_\mathrm{WD} = 1.4 M_\mathrm{\odot}$ and a
recurrence period of 22 years. 
%
%
Conversely, the nova models together with the above accretion rate of
10\% of the RG mass loss quantitatively constrain the mass loss of the
RG to values around $10^{-7} M_\mathrm{\odot}$yr$^{-1}$.  This is yet
another, completely independent, estimate of the mass loss rate of the
RG in RS Oph. It is more at the upper limit of the values given
by~\citet{1990ApJ...349..313S} for RG in classical symbiotics. A much
higher estimate of $10^{-5} M_\mathrm{\odot}$yr$^{-1}$ was derived
by~\citet{shore-et-al:96}. A more recent analysis indicates the rate
could be a factor of about 10 lower \citep{Shore:2008} and should only
be taken as an upper limit.  However, evolutionary models of single
stars predict values greater than $10^{-7} M_\mathrm{\odot}$yr$^{-1}$
for only a very short time on the RG
branch~\citep{1988A&AS...76..411M}. At the moment, we are not able to
resolve this spread in the estimates.
\begin{figure}[tp]
\centerline{\includegraphics[width=8.5cm]{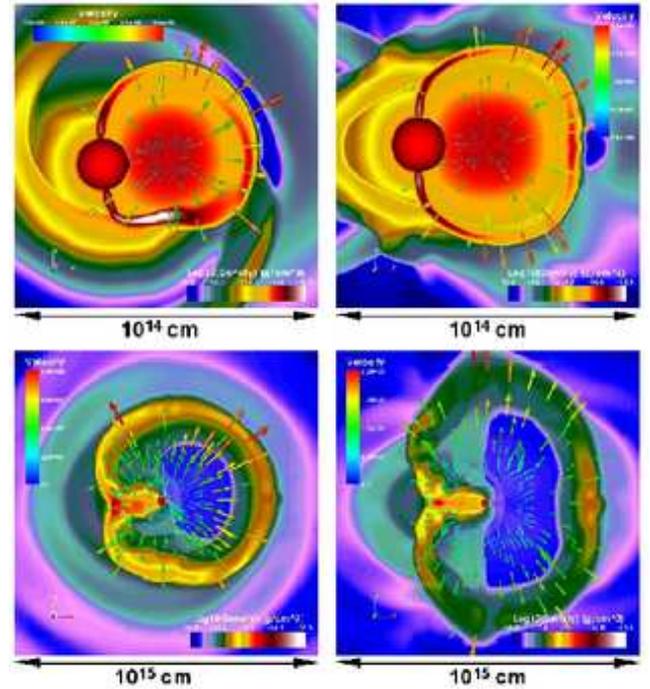}}
\caption{Density structure of the nova remnant. Shown are density
  (logarithmic scale, g/cm3) and velocity (cm/s) in the orbital plane
  (xy-plane, left) and in a plane perpendicular to the orbit
  (yz-plane, right) at 29 hours (top panels) and 21 days (bottom
  panels) after explosion. The RG is shown in red, the WD in blue.}
\label{fig:density_explosion}
\end{figure}
\subsection{Outburst: bipolar blast due to density stratification}
Launching the ejecta of a nova explosion into this complex
circumstellar environment naturally results in a strongly asymmetric
shock front, visible as a high density shell in
Fig~\ref{fig:density_explosion}. The time evolution of the nova
remnant is shown in a Supplementary Movie. The explosion was simulated
by instant injection of an energy and mass of $2.2 \cdot 10^{43}$ ergs
and $2 \cdot 10^{-7} M_\mathrm{\odot}$, respectively, at the position
of the WD. The assumed explosion mass is on the lower end of published
estimates for the ejected mass~\citep{sokoloski-et-al:06,
  obrien-et-al:92}, which range from $10^{-7} M_\mathrm{\odot}$ to
$10^{-6} M_\mathrm{\odot}$, but is comparable to the $2.2 \cdot
10^{-7} M_\mathrm{\odot}$ obtained from the self-consistently computed
accretion phase. An isothermal equation of state with $\gamma =1.1$ is
used, because inclusion of detailed cooling as in recent 1D
models~\citep{vaytet-et-al:07} is currently beyond reach in 3D
simulations.

The initial velocities of the modeled ejecta are $\sim 3500$~km/s,
within the range of observed values~\citep{bode-et-al:06,
  das-et-al:06, 2006MNRAS.373L..75E}. A larger extension of the shock
front in poleward directions, where pre-outburst densities are
smallest, is consistent with observations~\citep{obrien-et-al:07,
  chesneau-et-al:07}. In the orbital plane, the shock front displays a
roughly circular shape for about 2/3 of its arc. Toward the RG and the
former accretion wake, where densities are particularly high, the
shock front advances more slowly.

\begin{figure}[tp]
\centerline{ \includegraphics[width=9.cm,height=4.8cm]{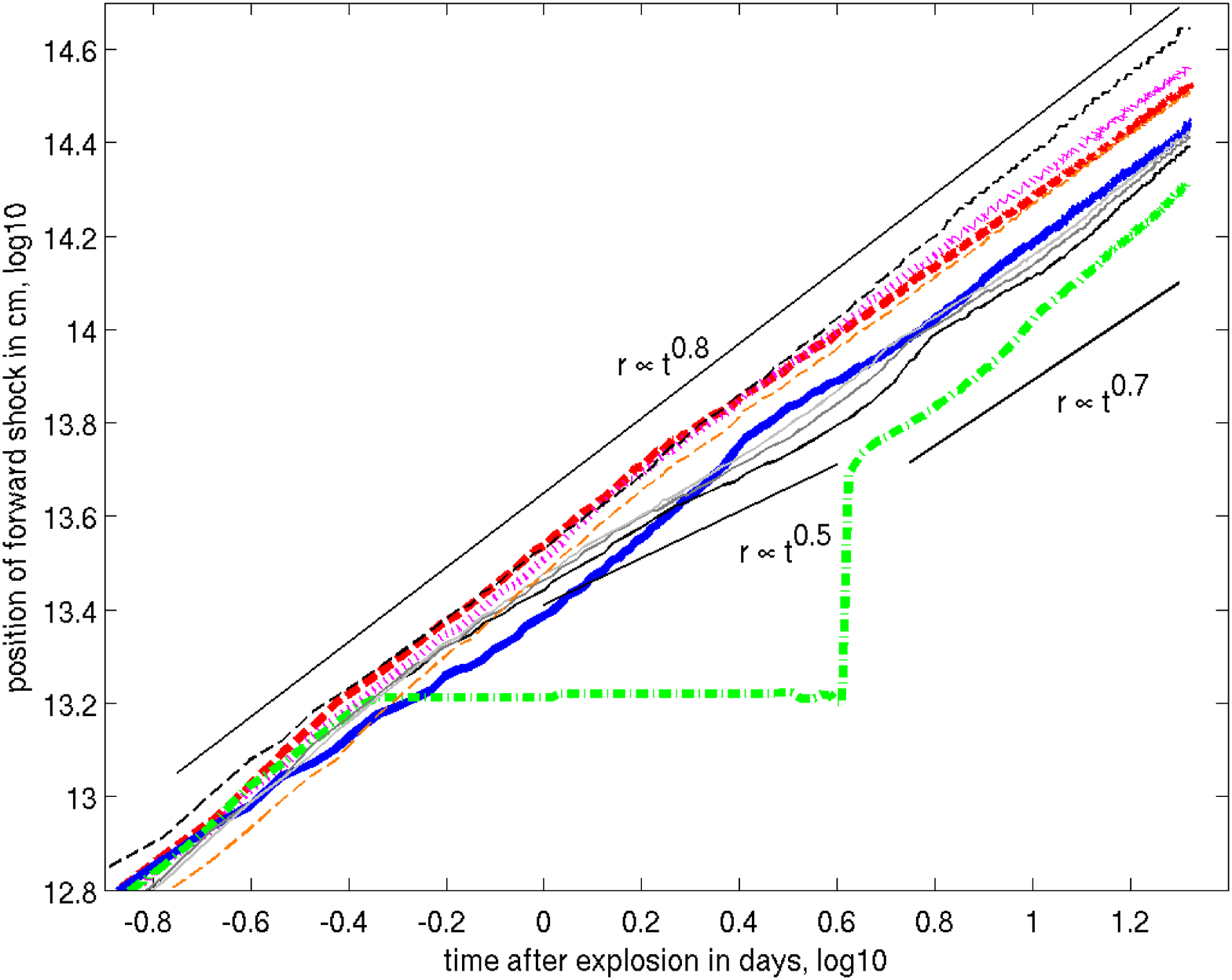}}
\caption{Time evolution of outer shock position relative to the WD.
  Bold colored lines belong to directions within the orbital plane
  (green: toward the RG or $+90^{\circ}$; blue: toward the trailing
  spiral of the WD or $0^{\circ}$; magenta: away from the RG or
  $-90^{\circ}$; red: in direction of the orbital motion or
  $-180^{\circ}$). The bold black line denotes the poleward direction.
  Thin lines belong to directions within the orbital plane, into the
  trailing accretion wake of the WD ($50^{\circ}$ to $60^{\circ}$,
  solid black to gray lines) and out of the system ($-35^{\circ}$,
  dashed orange line).}
\label{fig:explosion_velo_profile}
\end{figure}

Quantitative evaluation of the shock position as a function of time,
Fig~\ref{fig:explosion_velo_profile} yields shock velocities scaling
with time as $v_s \propto t^{-\alpha}$, with $0.2 \leq \alpha \leq0.5$
for most times and directions. Poleward values of $\alpha$ are
slightly smaller after about day 10, while $\alpha = 0.5$ for
directions toward the former accretion wake of the WD and times before
day 6 after the explosion. Values for $\alpha$ derived from
observational data of the 2006 nova outburst~\citep{bode-et-al:06,
  sokoloski-et-al:06, das-et-al:06} are in the same range up to day 6,
and are somewhat larger ($0.45 \le \alpha \le 0.79$) later on. {\it
  The present simulation suggests that the observed range of values is
  real and a consequence of different observational diagnostics
  probing different regions of the expanding shock front}. A clear
change in the velocity of the shock as a function of time is seen only
toward the former accretion wake.

There is no evident change in the expansion when the accumulated mass
equals the ejected mass of $2 \cdot 10^{-7} M_\mathrm{\odot}$, which
occurs after about eight to ten days. Observations taken much later
will show only mixed ejecta with the nova processed material
significantly diluted.
\vspace{-0.5cm}

\subsection{Shrinking of the binary orbit}
The orbit shrinks at a rate of $\dot{a}/a \approx -2.8 \cdot 10^{-8}$
per year, or 3\% per million years. This finding is in line with
previous suggestions~\citep{hachisu-et-al:99, 2005A&A...441..589J}.
The absolute losses of mass and angular momentum from the system,
$\dot{M}_\mathrm{S}$ and $\dot{J}_\mathrm{S}$, scale linearly with the
RG mass loss. For a RG mass loss rate of $10^{-7} M_\mathrm{\odot}$
per year, the fractional losses per year are $\dot{M}_\mathrm{S} /
M_\mathrm{S} = -2.2 \cdot 10^{-8}$ and $\dot{J}_\mathrm{S} /
J_\mathrm{S} = -4.4 \cdot 10^{-8}$, respectively, implying a
dimensionless specific systemic angular momentum loss of
$(\dot{J}_\mathrm{S} / J_\mathrm{S}) / (\dot{M}_\mathrm{S} /
M_\mathrm{S}) = 2$.  This time scale does not become significantly
larger if the effect of the nova blast on orbital parameters is taken
into account, provided not much more mass is ejected than was accreted
during the 22~years.
\section{Summary and conclusions}
Our 3D simulations of the quiescent phase result in a density
distribution of the circumstellar matter which is strongly stratified
in poleward directions, leading to an expansion of the nova remnant
that is roughly twice as fast in the poleward directions than within
the orbital plane. With a reasonable RG mass loss rate of about
$10^{-7} M_\mathrm{\odot}$yr$^{-1}$, the computed accretion rate of
about 10\% of the RG mass loss rate results in accretion rate of
$10^{-8} M_\mathrm{\odot}$yr$^{-1}$ for the WD. Inserting this value
into the multi-cycle nova models by~\cite{yaron-et-al:05} predicts on
a qualitative level -- though not in all quantitative details --
correctly the recurrence time, amplitude, and fastness of the RS~Oph
nova. Our simulations further predict shrinking of the binary orbit at
a rate of about 3\% per million years, which would improve conditions
for enhanced mass transfer. According to nova models
by~\cite{yaron-et-al:05}, enhanced mass loss should lead to shorter
nova cycles and favour net mass gain over multiple nova cycles, a
necessary condition for the evolution toward an SN Ia. On the other
hand, the above RG mass loss suggests that the mass and thus the Roche
volume of the RG will drastically shrink on a time scale of $\approx
10^6$ years, depending on the still not fixed mass of the star. The
system then would likely enter a common envelope phase and produce a
narrow WD-WD system.  Further angular momentum losses by gravitational
wave emission will drive the merger of the two WDs, an event which
should be detectable with present GW detectors, e.g.  Virgo and LIGO.
Depending on what the result of the merger is, an accretion induced
collapse or an SN~Ia will end the life of the system RS~Oph.
\begin{acknowledgements}
  We thank Dr. J. Favre from the Swiss Centre of Scientific Computing
  (CSCS) for graphics support and production of the movie accompanying
  this paper. Computations were performed at CSCS, at CINECA (Italy),
  and at the computing centre of ETH Zurich. The authors would like
  thank the staff of all supercomputing centres for substantial
  support. R.W. acknowledges financial support from the EGO Fellowship
  (VESF) and the INFN-Sezione Pisa, as well as the hospitality of the
  Max Planck Institute for Astrophysics, Garching, Germany, where part
  of the work was done.  SNS thanks J. Jos\'e, M. Bode, A.  Evans, and
  S. Starrfield for valuable discussions.
\end{acknowledgements}
\vspace{-0.5cm}
\bibliographystyle{aa} 
\bibliography{RsOphLetter.bib} 
\end{document}